\documentstyle[12pt]{article}
\newcommand{\bold}[1]{\mbox{\boldmath $#1$}} 

\newcommand{\mywarn}{******MY WARNING****** : }

\newcommand{\cref}{$\clubsuit$ \marginpar{$\clubsuit$ {\em Check ref.!}} 
\typeout{\mywarn Check reference(s)!}} 

\def\etc{ {\it etc}}

\def\ie{{\em i.e., }}
\def\d{\dagger}
\def\pbar{\bar{p}}
\def\nbar{\bar{n}}
\def\w {{\omega}}
\def\A {{{\cal A}}}
\def\B {{{\cal B}}}
\def\D {{{\cal D}}}
\def\bin#1#2{\left(\negthinspace\begin{array}{c}#1\\#2\end{array}\right)}
\def\mat#1#2#3#4{\left(\negthinspace\begin{array}{cc}
#1&#2\\#3&#4\end{array}\right)}
\def\be{\begin{equation}}
\def\ee{\end{equation}}
\def\br{\begin{eqnarray}}
\def\er{\end{eqnarray}}
\def\brn{\begin{eqnarray*}}
\def\ern{\end{eqnarray*}}
\def\rf#1{{(\ref{#1})}}
\def\ket#1{|#1 \rangle}
\def\bra#1{\langle #1|}
\def\Ket#1{||#1 \rangle}
\def\Bra#1{\langle #1||}
\begin{document}
\title{Relativistic RPA for Isobaric Analogue and Gamow-Teller
Resonances in Closed Shell Nuclei}
\author{C. De Conti and A. P. Gale\~{a}o  \\
{\it Instituto de F\'{\i}sica Te\'{o}rica,
 Universidade Estadual Paulista} \\
{\it Rua Pamplona 145,
 01405.900 S\~{a}o Paulo, SP,
 Brazil}\\
and \\ F. Krmpoti\'{c} \\
{\it Departamento de F\'{\i}sica, Facultad de Ciencias Exactas} \\
{\it Universidad Nacional de La Plata} \\ {\it C.C. 67
1900 La Plata,
Argentina}}
\maketitle
\begin{abstract}
We perform a self-consistent relativistic RPA calculation for the isobaric 
analogue and Gamow-Teller resonances based on relativistic mean field theory 
results for the ground states of $^{48}$Ca, $^{90}$Zr and $^{208}$Pb. We use
the  parameter set NL1 for the $\sigma$, $\omega$ and $\rho$ mesons, and
experimental values for the pion and nucleon. An extra parameter, related
to the intensity of the contact term in the pion-exchange interaction,
is crucial to reproduce the latter resonances.
\end{abstract}
\newpage
In recent years the relativistic mean field theory (RMFT) has  been successfully
applied to account for both:
i) the bulk features of nuclear matter (saturation, equation of state,\etc.) 
and
ii) the ground state properties of finite nuclei, including unstable ones up
to the nucleon drip lines \cite{Men96}. (For reviews, see refs.
\cite{Ser86,Wal95,Rin96,Ser97}.)
A nice feature of the model is that  the same restricted set of parameters
can be used for all these systems. Even though some excited nuclear states,
including certain  giant resonances, have also been calculated in the
relativistic random phase approximation (RRPA) based on RMFT, this has
never been done for charge-exchange excitations.   The aim of the present
work is to explore to what extent this relativistic model, whose parameters
have been fitted to ground state properties, can describe the charge-exchange
collective states, such as the isobaric analogue state (IAS) and the
Gamow-Teller resonance (GTR).

Both IAS and GTR have been extensively calculated  in the context of
 traditional, \ie nonrelativistic, nuclear structure theory, usually in the 
random phase approximation (RPA) or extensions of it, making
use of, not only completely phenomenological interactions, but also 
more realistic ones based on meson exchange. While the IAS is easily 
reproduced, the GTR is more problematic, especially with respect to 
its strength that turns out to be quenched to $\sim$ 60\%  
of the Ikeda sum-rule value. (For a review of GTR, 
see ref. \cite{Ost92}.) Two mechanisms have been proposed for
the absorption of the unseen strength: i) the excitation of the 
$\Delta$ resonance, which has the same quantum numbers as the GTR, 
or ii) the influence of 2p-2h and higher excitations,
not included in RPA. We shall not dwell on such finer points 
in this paper, since this would require, anyway, a more sophisticated 
treatment than the simple version of RRPA we are using, but rather 
will concentrate on the quality of the reproduction of excitation 
energy and strength as compared with similar nonrelativistic calculations
\cite{Kre81,Col94}.

Relativistic mean field theory (RMFT) always takes as a starting point some 
version of an effective quantum field theory describing a system of interacting 
hadrons, coupled also to the electromagnetic (or to the electroweak) field, 
generally referred to as quantum hadrodynamics (QHD)
\cite{Ser86,Wal95,Rin96,Ser97}. 
%
Historically, QHD was introduced as a renormalizable quantum field theory,
which severely limited the number of admissible terms in its
Lagrangian \cite{Ser86}. However, despite its many successes, it
had difficulties with its renormalization program.
Thus, in the last years, it began to be formulated
as an effective, nonrenormalizable quantum field theory
representing the low-energy limit of the fundamental theory of strong
interactions, namely, quantum chromodynamics (QCD). Consequently,
all terms compatible with the symmetries of QCD were now admissible, and one
had to find appropriate expansion parameters and a criterion of {\em
naturalness} to select the most important ones. Even thus, the number of terms
increased significantly. Their coupling constants were adjusted to a judicious
selection of nuclear properties that should be well reproduced at the mean
field level and, while this was  successful in the sense that they turned out
to be {\em natural}, on the other hand there are clear indications that their
full set is underdetermined \cite{Ser97}.
Since our main purpose here is not to test the foundations of QHD, but rather
to assess its ability to reproduce some new nuclear properties, and we want to
keep the number of free parameters to a minimum, we will include only the most
important terms, and work with the following nonrenormalizable QHD Lagrangian 
density:
%
\begin{eqnarray}  
{\cal L} & = & \bar{\psi} (i\gamma_{\mu} \partial^{\mu} - M)\psi \nonumber \\      
         &   & \mbox{} + \frac{1}{2} \partial_{\mu} \sigma \partial^{\mu} \sigma 
               - \frac{1}{2} {m_{\sigma}}^{2} \sigma^{2} - \frac{1}{3} g_{2}           
               \sigma^{3} - \frac{1}{4} g_{3} \sigma^{4} - g_{\sigma} \bar{\psi} 
               \psi \sigma \nonumber \\
         &   & \mbox{} - \frac{1}{4} \Omega_{\mu\nu} \Omega^{\mu\nu} 
               + \frac{1}{2} {m_{\omega}}^{2} \omega_{\mu} \omega^{\mu} 
               - g_{\omega} \bar{\psi} \gamma_{\mu} \psi \omega^{\mu} \nonumber
\\
         &   & \mbox{} + \frac{1}{2} \partial_{\mu} \bold{\pi} \cdot             
               \partial^{\mu} \bold{\pi} - \frac{1}{2} {m_{\pi}}^{2} \bold{\pi}  
               \cdot \bold{\pi} - \frac{f_{\pi}}{m_{\pi}} \bar{\psi} \gamma_{5}  
               \gamma_{\mu} \bold{\tau} \psi \cdot \partial^{\mu} \bold{\pi}
\nonumber \\ 
         &   & \mbox{} - \frac{1}{4} {\bold{R}}_{\mu\nu} \cdot \bold{R}^{\mu\nu} 
               + \frac{1}{2} {m_{\rho}}^{2} \bold{\rho}_{\mu} \cdot              
               \bold{\rho}^{\mu} - \frac{g_{\rho}}{2} \bar{\psi} \gamma_{\mu}    
               \bold{\tau} \psi \cdot \bold{\rho}^{\mu} \nonumber \\
         &   & \mbox{} - \frac{1}{4} F_{\mu\nu} F^{\mu\nu} - e \bar{\psi}        
               \gamma_{\mu} \frac{1+\tau_{3}}{2} \psi A^{\mu},
\label{1}
\end{eqnarray}
in standard notation ($\hbar = c = 1$), where $\psi$ is the spin-$\frac{1}{2}$, 
isospin-$\frac{1}{2}$ nucleon field, $\sigma$, $\omega^{\mu}$, $\bold{\pi}$ and
$\bold{\rho}^{\mu}$ denote the meson fields, and $A^{\mu}$
is  the electromagnetic field \cite{Wal95}.
The field tensors
for the vector particles read
\begin{eqnarray}
\Omega^{\mu\nu} & = & \partial^{\mu} \omega^{\nu}-\partial^{\nu}\omega^{\mu},
\nonumber\\
\bold{R}^{\mu\nu} & = & \partial^{\mu} \bold{\rho}^{\nu} - \partial^{\nu}
\bold{\rho}^{\mu} - g_{\rho} \bold{\rho}^{\mu} \times \bold{\rho}^{\nu},
\label{2}\\
F^{\mu\nu} & = & \partial^{\mu} A^{\nu} - \partial^{\nu} A^{\mu}.
\nonumber
\end{eqnarray}

To reproduce the experimental values of the nuclear incompressibility and
surface
diffuseness in RMFT, it has been found important \cite{Bog77} to allow the
sigma meson to self-interact. This is done through the cubic and quartic terms,
in $\sigma$, appearing in eq.~\rf{1}.
All couplings are taken to be of the direct,
nonderivative kind, except in the pion-nucleon case, where both the
phenomenology
and theoretical reasons (related to chiral symmetry) definitively favor the
pseudovector coupling. It is true that phenomenology would also recommend that a 
certain amount of derivative coupling should be mixed in for the vector mesons 
to the
nucleon \cite{Mac89}. Yet, for simplicity, we followed the general trend in RMFT
calculations \cite{Men96,Gam90,Rei86} and did not include such terms in
eq.~\rf{1}. For the
omega meson, this can be justified since that term would be rather small 
\cite{Bou87},  but 
for the rho meson, while unimportant for the nuclear ground state, the 
corresponding term would give a sizable contribution to excitations of the 
Gamow-Teller type. 
Finally, the photon is minimally coupled 
to the proton  to take care of the important 
Coulomb interaction inside nuclei, but not to the charged mesons since this 
would be of much less relevance for the properties of interest here and is 
 hence ignored in this model Lagrangian.

The nucleon mass \footnote{The proton mass is used here.}, $M$, the proton 
charge, $e$, the pion mass, $m_{\pi}$, and
coupling constant, $f_{\pi}$, are fixed at their experimental values. Thus, the 
masses of the remaining mesons included in eq. \rf{1}, $m_{\sigma}$,
$m_{\omega}$ and $m_{\rho}$, their coupling constants, $g_{\sigma}$, 
$g_{\omega}$ and $g_{\rho}$, and the self-interaction strengths for the sigma 
meson, $g_{2}$ and $g_{3}$, give a total of only 8 free parameters for this 
model.

RMFT results from an approximation scheme to solve the Euler-Lagrange 
equations derived from ${\cal L}$, which consists in taking 
advantage of the high values of the  densities inside nuclei to  replace the 
meson and photon fields by classical fields satisfying these equations 
with the source terms replaced by their expectation values in the nuclear 
ground state of interest. This is equivalent to compute  the meson
self-energies in the Hartree approximation 
%
disregarding the negative-energy, i.e.,  antiparticle,  states, which is
sometimes referred to as the no-sea approximation.
%

The RMFT equations simplify considerably if one takes advantage of known 
symmetries of the nuclear ground state. Firstly,
as it always has a
definite parity, the pion, due to its pseudoscalar nature,  
completely disappears from these equations. 
Secondly, the definite charge of the nucleus eliminates all but the third
component of the isovector mesons. Also, we 
are interested here 
only in spherical nuclei, and their rotational invariance implies
that the spatial components of the currents in the source terms 
vanish in RMFT. 
So, the
corresponding components of the vector fields can be ignored. Finally, we 
are looking only for stationary solutions in the nuclear rest frame. 
Thus,
the meson and photon fields are taken as time-independent and the 
nucleon field has merely  time-dependent phase factors, i.e., 
\begin{equation}  
\psi(x_\mu) = \sum_{\alpha} a_\alpha  {\cal U}_{\alpha} (\bold{r})
\exp ( - i E_{\alpha} t),
\label{3}
\end{equation}
where the no-sea approximation has been made, and $a_{\alpha}$ is the
annihilation operator in the  particle state
with (positive) energy $E_\alpha$.
With these simplifications, the RMFT equations become
\begin{eqnarray}
\left[- i \bold{\alpha} \cdot \nabla  + \beta \left( M + {V_s}(r) \right)
+ {V_v}(r) \right] {\cal U}_\alpha &=& E_\alpha {\cal U}_\alpha,
\label{4}\end{eqnarray}
and
\begin{eqnarray}
\nonumber \\
\left( - \nabla^2 + {m_\sigma}^2 \right) \sigma &=& - g_\sigma 
{\rho_s}(r) - g_2 \sigma^2 - g_3 \sigma^3,
\nonumber \\
\left( - \nabla^2 + {m_\omega}^2 \right) \omega^0 &=& g_\omega
{\rho_v}(r),  \label{5}\\
\left( - \nabla^2 + {m_\rho}^2 \right) \rho_3^0 &=& g_\rho {\rho_3}(r),
\nonumber\\
- \nabla^2 A^0 &=& e {\rho_p}(r),
\nonumber
\end{eqnarray} 
where
\begin{eqnarray}
V_s &=& g_\sigma \sigma,~~
V_v = g_\omega \omega^0 + g_\rho \tau_3 \rho_3^0 + e \frac{1+\tau_3}{2}
A^0,
\label{6}
\end{eqnarray}
are the scalar and vector potentials, and
\begin{eqnarray}
\rho_s &=& \left \langle \bar{\psi} \psi \right \rangle,~
\rho_v =\left \langle \bar{\psi} \gamma^0 \psi \right \rangle,~
\rho_3 =\left \langle \bar{\psi} \gamma^0 \tau_3 \psi \right \rangle,~
\rho_p =\left \langle \bar{\psi} \gamma^0 \frac{1+\tau_3}{2} \psi \right
\rangle,
\label{7}
\end{eqnarray}
the scalar, vector, isovector (third
component) and proton densities.
The expectation values are taken in the nuclear ground state, constructed 
by putting $Z$ protons and $N$ neutrons, ($Z+N=A$), in the lowest 
particle states with $\tau_3$ equal to $+1$ and $-1$, respectively, 
obtained by solving the Dirac equation \rf{4}. This yields
the following results for the densities
\begin{eqnarray}
\rho_s &=& \sum_{\alpha=1}^{A} \bar{\cal U}_\alpha
{\cal U}_\alpha,~
\rho_v =\sum_{\alpha=1}^{A} \bar{\cal U}_\alpha
\gamma^0 {\cal U}_\alpha,
\nonumber\\
\rho_3 &=&\sum_{\alpha=1}^{A} \bar{\cal U}_\alpha
\gamma^0 \tau_3 {\cal U}_\alpha,~
\rho_p = \sum_{\alpha=1}^{A} \bar{\cal U}_\alpha
\gamma^0 \frac{1+\tau_3}{2} {\cal U}_\alpha.
\label{8} \end{eqnarray}
The set of eqs. \rf{4}--\rf{8}
should be solved self-consistently.

%
Alternatively, the RMFT equations could have been obtained by extremization of
an energy functional built from eq. \rf{1} and involving only valence nucleons
and classical photon and meson fields. This density functional approach is
sometimes preferred when working strictly within the context of an effective
field theory, since it becomes clear in this case that much of the effects of
vacuum dynamics and of certain many-body correlations beyond the Hartree level
are absorbed in the renormalized coupling constants in the Lagrangian, when
fitted to experimental data \cite{Ser97,Fur95}. 
%

The RRPA equations have been derived several times in the literature  
and applied, both to nuclear matter \cite{Kur85}, and to finite nuclei
\cite{Blu88}--\cite{Ma97}. For our purpose here, they can be obtained
from the  equations of motion formalism of Rowe \cite{Row70}.
One introduces charge-exchange excitation operators of the form 
\cite{Row75,Lan80,Bau88}
\begin{eqnarray}
O^\dagger_{J\lambda}=\sum_{p \bar{n}} X_{p \bar{n}}^{J\lambda}
\left({a_p}^\dagger a_{\bar{n}}\right)_J
-\sum_{n \bar{p}} Y_{n \bar{p}}^{J\lambda}
\left({a_{\bar{p}}}^\dagger a_n\right)_J,
\label{9} \end{eqnarray}
where $p$ and $\bar{p}$ ($n$ and $\bar{n}$) label unoccupied and occupied 
proton (neutron) positive-energy, single-particle RMFT states. 
$J$ is the total angular momentum, and $\lambda$ runs from $1$ to $\pi+\nu$,
with $\pi$ and $\nu$ being the number of
$p\nbar$ and $n\pbar$ excitations, {\it i.e.}, those having 
\begin{math} 
\sum_{p \bar{n}} {| X_{p \bar{n}}^{J\lambda} |}^2 -
\sum_{n \bar{p}} {| Y_{n \bar{p}}^{J\lambda} |}^2
\end{math}
equal to $+1$ and $-1$, respectively.
Following the standard procedure one arrives at the RRPA equations
\be
\mat{\A^J}{\B^J}{\B^{J\d}}{\D^J}
\bin{X^{J\lambda}}{Y^{J\lambda}}
=\w_{J\lambda}\bin{X^{J\lambda}}{-Y^{J\lambda}}.
\label{10}\ee
The submatrices are given in
terms of the residual interaction $V$ by
\begin{eqnarray}
\A_{p\nbar,p'\nbar'}^J &=& (E_p-E_{\nbar})\delta_{pp'}\delta_{\nbar\nbar'} + 
\bra{(p\nbar^{-1})_J}V\ket{(p'\nbar'^{-1})_J},
\nonumber\\
\D_{n\pbar,n'\pbar'}^J&=&(E_n-E_{\pbar})\delta_{nn'}\delta_{\pbar\pbar} + 
\bra{(n\pbar^{-1})_J}V\ket{(n'\pbar'^{-1})_J},
\label{11}\\
\B_{p\nbar,n'\pbar'}^J&=& 
(-)^{J+M_J} \bra{(p\nbar^{-1})_{JM_J} (n'\pbar'^{-1})_{J,-M_J}}V\ket{}.
\nonumber \end{eqnarray}
The excitation energies $E^+_{J\lambda_+}$ ($\lambda_+=1,\cdots \pi$)
in the odd-odd $(N-1,Z+1)$ nucleus (measured from the target ground state)
are the highest $\pi$ eigenvalues $\omega_{J\lambda}$ in \rf{10}. In the
same way the excitation energies $E^-_{J\lambda_-}$ ($\lambda_-=1,\cdots \nu$)
in the $(N+1,Z-1)$ nucleus are the remaining  $\nu$ eigenvalues
$\omega_{J\lambda}$ taken with opposite sign.
The corresponding total transition strengths are:
\begin{eqnarray}
S^+_J&=& \sum_{\lambda_+}\left|\sum_{p\bar{n}}X_{p \bar{n}}^{J\lambda_+*}
 \Bra{p} w_J\Ket{\bar{n}} +\sum_{n \bar{p}} Y_{n \bar{p}}^{J\lambda_+*}
\Bra{\bar{p}}w_J\Ket{n} \right|^2,
\nonumber \\
S^-_J&=& \sum_{\lambda_-}\left|
\sum_{n \bar{p}} Y_{n \bar{p}}^{J\lambda_-}
\Bra{\bar{p}}w_J\Ket{n}
+\sum_{p\bar{n}}X_{p \bar{n}}^{J\lambda_-}
\Bra{p}w_J\Ket{\bar{n}}\right|^2,
\label{12}\end{eqnarray}
where
$w_0=1$ and $w_1=\bold{\sigma}$.
They fulfill the well known sum rules:
\be
S^+_J-S^-_J=(2J+1)(N-Z).
\label{13}\ee

For a self-consistent calculation, $V$ must be obtained from the same Lagrangian,  
given in eq. (\ref{1}), which was used for the mean field part. Also, to be
compatible with RMFT, one must consider only the direct terms in eq.
(\ref{11}), leading, in fact, to a ring approximation. 
Hence,
only the isovector mesons contribute to isovector excitations like IAS and GTR. 
Furthermore, when the instantaneous approximation is made,
the interaction, $V=V_\pi+V_\rho$, reads
\begin{eqnarray}
V_\pi(1,2) &=& - \left(\frac{f_{\pi}}{m_{\pi}}\right)^2
\bold{\tau}_1\cdot\bold{\tau}_2 \; 
(\bold{\sigma}_1\cdot\nabla_1 \; \bold{\sigma}_2\cdot\nabla_2) Y(m_\pi,r),
\nonumber\\
V_\rho(1,2) &=& {g_\rho}^2 \bold{\tau}_1\cdot\bold{\tau}_2 \; 
( 1 - \bold{\alpha}_1\cdot\bold{\alpha}_2 ) Y(m_\rho,r),
\label{14}\end{eqnarray}
with $r=\left|\bold{r}_1-\bold{r}_2\right|$ and $Y(m,r)= \exp(-mr)/(4\pi r)$.

We performed our numerical calculations taking $^{48}$Ca, $^{90}$Zr and 
$^{208}$Pb as the target nuclei. 
To get a discretized set of single-particle RMFT states, we solved 
eqs. \rf{4}--\rf{8} by expanding the different fields in truncated
harmonic oscillator bases, following the method of P. Ring and 
collaborators as described in ref. \cite{Gam90}, where the details 
can be found. For the RMFT parameters we chose the set 
NL1, which is widely used for nuclear ground state calculations 
\cite{Men96,Gam90,Rei86}. The RRPA equations were solved for
$J^{\pi}=0^+$ and $1^+$ states in a  
model-space including only $0\hbar \Omega$ and $2\hbar \Omega$ 
excitations.
Furthermore, since the continuum is not well represented 
by an expansion in a harmonic oscillator basis, we tried to avoid
it as much as possible. To this end, we included only 
those RMFT single-particle states that are bound at least for 
neutrons. (Of course, for consistency, we had to include the 
corresponding states for protons, even when unbound.) That this 
model-space is not too restricted is attested by the fact that, 
in all the cases considered here, the
sum rule \rf{13} was obeyed at the level of 99.5\% (92.8\% ) or 
better for the IAS (GTR).

Our results for the IAS of $^{208}$Pb give an excitation energy 
(always measured from the target ground state) of 18.6 MeV 
and a strength of 99\% of the sum-rule prediction of $(N-Z)$, 
which compare well with the experimental values of 
18.8 MeV and $\sim$ 100\%, as quoted in \cite{Aki95}. 
These results are of the same quality as those of Col\`o et al. 
\cite{Col94}, who perform a nonrelativistic calculation 
that is also self-consistent, in the sense that they use the same
interaction (Skyrme force)
for the excitations 
as for the target ground state and single-particle
basis.

Since the  GTR's are rather broad, having experimental widths
of $\sim 4$ MeV, the theoretical results we report here
correspond to a smoothed strength function, constructed by
replacing the spikes in the RRPA strength distribution by Lorentzian peaks of 
conveniently chosen width. The excitation energy and strength of the GTR are 
then obtained by fitting a Lorentzian to the main peak in this strength 
function.
 When the interaction \rf{14} is used we get that the excitation energy
and the strength of the GTR in $^{208}$Pb are, respectively, $\sim 6$ MeV
and $56\%$ of the Ikeda sum-rule, while the corresponding experimental
values are $19.2$ MeV and $60-70\%$ \cite{Aki95}.
Therefore, as it stands,  the relativistic calculation would
underestimate the GTR by more than $10$ MeV, putting it much lower than
the IAS, in blatant opposition to experiment. 
The nonrelativistic calculation of Col\`{o} et al. does much better,
though it overestimates the excitation energy by $2-3$ MeV.

It is still possible, however, 
to improve the GTR results in our
calculation, without losing the self-consistency between
the excitation and the mean field parts.  In fact, when dealing with the
pion contribution, we might follow the general practice and
eliminate the contact term, which comes from
the derivative (pseudovector)  coupling and is thought to be 
suppressed by the short-range correlations between nucleons. 
(As such it should disappear when a form factor is used to
account for finite nucleon and meson sizes.)
More, to reproduce the energetics of the GTR within
a nonrelativistic
calculation based on $\pi$- and $\rho$- exchanges,
Krewald et al. \cite{Kre81} were forced to
add a zero-range Landau-Migdal interaction of adjustable strength.
In our case this would correspond to  gauge the contact term in the
bare $\pi$-exchange interaction, given by \rf{14}, or equivalently
to add the term
\begin{equation}
\delta V_{\pi}(1,2) =  g' \left( \frac{f_{\pi}}{m_{\pi}} \right)^2 
\bold{\tau}_1 \cdot \bold{\tau}_2 \;
\bold{\sigma}_1 \cdot \bold{\sigma}_2 \;
\delta (\bold{r}_1 - \bold{r}_2),
\label{15}\end{equation}
which for $g'=1/3$ completely cancels the contact part in \rf{14}.
This term was found to play an important role in the RRPA
calculation of unnatural parity states in $^{16}$O made by Blunden and
McCorquodale \cite{Blu88}, who used the purely phenomenological value
$g'=0.7$, in analogy to nonrelativistic calculations.
Since we are neglecting exchange contributions, $\delta V_{\pi}$ has no
effect on the IAS, but is of overwhelming importance for the GTR,
as illustrated in Fig.~\ref{pbgt} for the case of $^{208}$Pb. One sees
that the theoretically  justified coupling $g'=1/3$ leads to better results
than the bare interaction ($g'=0$).
Yet, the agreement with experiment is only achieved with $g'\cong 0.7$.
Thus, we found it necessary to incorporate $\delta V_{\pi}$, with $g'=0.7$,
into the residual interaction.

The results obtained in our calculation, with this choice of residual 
interaction, for the IAS and GTR, excited from the closed-shell nuclei 
$^{48}$Ca,
$^{90}$Zr and $^{208}$Pb, are summarized in Table~\ref{gtia}. One can see that
they are of the same quality as those obtained in similar nonrelativistic 
calculations \cite{Col94}. The agreement with experiment is very good
for the excitation
energies of both resonances, but less so for the strengths, specially for
the GTR.
This was already expected,
since our model
does not include either 2p-2h or
$\Delta$-h excitations, and therefore
is unable to reproduce the quenching of the GTR.
For similar reasons,
and also because we do not treat the 
continuum properly, we get very little strength at higher excitation energies 
beyond the GTR peak. 
In the case of $^{90}$Zr, for instance, our calculated 
strength function is decaying well below 0.45 MeV$^{-1}$ 
already for $\sim 25$ 
MeV excitation, while the experimental one \cite{Wak97} stays 
at about this 
value up to $\sim 60$ MeV.

We conclude that both the IAS and the GTR excited from closed shell nuclei can 
be well reproduced  in the context of quantum hadrodynamics, with
the ground state parametrization for the $\sigma$, $\omega$ and $\rho$ mesons,
and neglecting the exchange contributions.
For the GTR, however, the pion  with pseudovector coupling to the nucleon
also must be included, and it is essential to use for the strength
$g'$ of the contact term $\delta V_\pi$ a value close to that
used in similar nonrelativistic calculations.
An alternative approach, to account for the energies of the GTR's,
might be the mixed coupling of the rho meson to the
nucleon. Yet, such a parametrization for ground state properties is available
only in the Hartree-Fock approximation \cite{Bou87}.

The authors wish to thank Peter Ring for enlightening discussions on the general 
topic of RMFT calculations and for giving them access to his spherical RMFT code 
and assistance in its usage.
The work of C.D.C. was supported by CAPES (Brazil). A.P.G. and F.K. acknowledge
partial financial support by ICTP (Trieste), CONICET (Argentina) and CLAF (Latin
America).
%
%

%
%
%
\begin{figure}[p] 
\vspace{5cm}
\caption{Theoretical Gamow-Teller strength distribution for the parent nucleus 
$^{208}$Pb obtained with an isovector interaction including a Landau-Migdal-type 
contact term of strength $g'$ = 0, 1/3 and 0.7. The spikes give the raw RRPA 
results and the continuous curve, the strength function smoothed out as 
explained in the text. The strength function for the resonance peak
extracted from experiment 
\protect\cite{Aki95} is drawn in dotted line. \label{pbgt} }
\end{figure}
%
%
%
\begin{table}[p] 
\caption{RRPA results for excitation energies and strengths of charge-exchange 
resonances calculated with isovector interaction including a Landau-Migdal-type 
contact term of intensity $g'=0.7$, as described in the text. Experimental 
values are given for comparison. \label{gtia}}
\begin{minipage}[t]{12cm}
   \begin{tabular}{lccccc} \hline \hline
   Parent      & \multicolumn{2}{c}{RRPA}  & & \multicolumn{2}{c}{Experiment} \\
                 \cline{2-3}                \cline{5-6}
   and         & Energy & Strength         & & Energy & Strength \\
   resonance   & [MeV]  & [\% of sum rule] & & [MeV]  & [\% of sum rule] \\ 
   \hline
    & & & & & \\
   $^{208}$Pb\footnote{Experimental values taken from Refs. \cite{Hor80,Aki95}.}
    & & & & & \\     
   IAS & 18.6 & 99    & & $18.83 \pm 0.02$ & $\sim 100$ \\
   GTR & 18.9 & 80    & & $19.2 \pm 0.2$   & 60 -- 70   \\
    & & & & & \\
   $^{90}$Zr\footnote{Experimental values taken from Refs. \cite{Bai80,Wak97}.}
    & & & & & \\ 
   IAS & 11.9 & 101.3 & & $12.0 \pm 0.2$   & $\sim 100$ \\
   GTR & 16.0 & 80    & & $15.6 \pm 0.3$   & 28         \\
    & & & & & \\
   $^{48}$Ca\footnote{Experimental values taken from Ref. \cite{And85}.}
    & & & & \\
   IAS & 7.48 & 100.4 & & 7.17             & $\sim 100$ \\
   GTR & 11.0 & 82    & & $\sim 10.5$      & 35         \\ \hline \hline
   \end{tabular}
\end{minipage}   
\end{table}
\end{document}